# Quantum Conductance Probing of Oxygen Vacancies in SrTiO$_3$ Epitaxial Thin Film Using Graphene

*Kyeong Tae Kang, Haeyong Kang, Jeongmin Park, Dongseok Suh\*, and Woo Seok Choi\**

K. T. Kang
Department of Physics, Sungkyunkwan University, Suwon 16419, Korea
Center for Integrated Nanostructure Physics, Institute for Basic Science (IBS), Suwon 16419, Korea

Dr. H. Kang
Department of Energy Sciences, Sungkyunkwan University, Suwon 16419, Korea

J. Park
Center for Integrated Nanostructure Physics, Institute for Basic Science (IBS), Suwon 16419, Korea
Department of Energy Sciences, Sungkyunkwan University, Suwon 16419, Korea

Prof. D. Suh
Department of Energy Sciences, Sungkyunkwan University, Suwon 16419, Korea
E-mail: energy.suh@skku.edu

Prof. W. S. Choi
Department of Physics, Sungkyunkwan University, Suwon 16419, Korea
E-mail: choiws@skku.edu

The quantum Hall conductance in monolayer graphene on an epitaxial SrTiO$_3$ (STO) thin film is studied to understand the role of oxygen vacancies in determining the dielectric properties of STO. As the gate voltage sweep range is gradually increased in our device, we observe systematic generation and annihilation of oxygen vacancies evidenced from the hysteretic conductance behavior in graphene. Furthermore, based on the experimentally observed linear scaling relation between the effective capacitance and the voltage sweep range, a simple model is constructed to manifest the relationship among the dielectric properties of STO with oxygen vacancies. The inherent quantum Hall conductance in graphene can be considered as a sensitive, robust, and non-invasive probe for understanding the electronic and ionic phenomena in complex transition metal oxides without impairing the oxide layer underneath.

Oxygen plays an indispensable role in determining the functional properties of transition metal oxides (TMOs), such as superconductivity, memristive behavior, and topotactic phase reversal.[1-6] In STO, a prototypical perovskite oxide, changes in both electronic and ionic behaviors were observed by introducing oxygen vacancies.[7] Oxygen vacancies in STO can be generated via the electric-field-induced redox process or the application of chemical pressure at high temperatures. They supply conducting electrons ($3d^1$) to normally insulating STO ($3d^0$), which can eventually form Cooper pairs at low temperatures (~1 K).[8] On the other hand, oxygen vacancies generated/redistributed via electric-field-induced redox reaction near the metal/oxide interface might further modify the metal/STO heterojunction and give rise to functional phenomena such as resistive switching.[1-2,6,9-13] Modulation of the local conductivity using a biased conductive atomic force microscopy tip and observation of gas bubbles beneath the anodic metal electrodes indicate the electroforming of oxygen vacancies which affect the transport property of STO critically.[10-11] More recently, it has also been shown that oxygen vacancies can be created just by depositing the active metal, directly observed by the transmission electron microscopy (TEM).[14] However, the role of the ionic motion in modifying the fundamental dielectric properties of STO near the heterojunction is yet to be understood, despite its important implications in the electronic applications.

In order to perceive the formation of oxygen vacancies and estimate their influence on the dielectric behavior of STO, we propose the 2D transport nature of graphene as a sensitive, robust, and non-invasive probe for examining the layer underneath. Among various characteristics of graphene, linear dispersion relation and the gapless feature give rise to a delicate electronic response to an external electric field. In addition, a universal quantum Hall effect in graphene is exclusively dependent on its carrier concentration. Therefore, the position and shape of the charge neutrality point (CNP) and the plateaus originating from the quantum Hall states can be considered as rigorous measures for assessing the electronic environment, which graphene is experiencing.[15]

The aforementioned approach is particularly interesting for probing the TMO layer. Owing to the carrier-sensitive transport behavior, graphene-ferroelectric oxide devices exhibit not only a hysteretic transport behavior but also a clear polarization-dependent switching of CNP polarity.[16-21] High-*k* oxides have also been studied to examine the effective screening of the charge carriers in graphene.[22-26] While the expected large enhancement of carrier mobility has not been observed in graphene near a highly dielectric environment,[27] the effective scaling of an electric field has been realized with a clear quantum Hall behavior.[28]

In this communication, we investigated the electrical transport properties of monolayer graphene on an epitaxial STO thin film and examined the redox activity at the graphene-STO interface. Our high quality graphene-STO device is free from noticeable defects that might act as charge trapping sites, manifested by the observation of clear quantum Hall effect. On top of the quantum Hall effect, we observed the development of hysteretic behavior and a systematic shift of the CNP when the gate voltage sweep range was increased above a certain value. These phenomena could be consistently understood in terms of gate-voltage-dependent creation and annihilation of oxygen vacancies due to the electric-field-induced redox activity in the oxide layer near the oxide-graphene interface. By analyzing the scalable quantum conductance, we further provide a quantitative estimate of the change in the dielectric properties of STO due to the oxygen vacancy formation.

The schematic geometry and the creation/annihilation of oxygen vacancies via the electric field in our graphene-STO device are shown in **Figure 1**. The device consists of a Nb:STO single crystal substrate, 90-nm-thick stoichiometric epitaxial STO thin film, and monolayer graphene, which serve as the bottom electrode, gate dielectric, and channel, respectively. Our epitaxial STO thin film and the transferred monolayer graphene are of high crystalline quality with minimized defects, clearly evidenced by the XRD $\theta$-$2\theta$ scan (Figure 1a) and atomic force microscopy.[7,28] Without oxygen vacancies, stoichiometric STO can be considered as a normal gate dielectric, and graphene will exhibit a typical transport behavior with voltage scaling (Figure 1b).[27-28] On the other hand, when charged oxygen vacancies are introduced near the surface of STO, graphene will experience an effective electric field, which is equivalent to hole doping (Figure 1c).

The gate-voltage ($V_G$)-dependent conductance of our graphene-STO device exhibits a systematic development of the hysteretic behavior, as shown in **Figure 2**a. For $V_G$ sweep range of < 1 V, no hysteresis is observed in the conductance, which is consistent with the previous result on graphene on a 300 nm $SrTiO_3$ epitaxial thin film.[28] However, for $V_G$ sweep range of ≥ 1 V, the hysteretic conductance develops. The $V_G$ sweep sequence is denoted by arrows and alphabets in Figure 2a for a large $V_G$ sweep range, which starts from 0 to positive $V_G$. Below a certain positive voltage, the graphene conductance shows a typical linearly increasing behavior. Then, when the gate voltage exceeds the threshold voltage ($V_{th} \approx$ 1 V), the source-drain current ($I_{SD}$) saturates. This marks the beginning of the hysteresis. When $V_G$ is decreased, $I_{SD}$ starts to decrease immediately and follows again a typical linear conductance behavior with shifted CNP to a positive voltage. As $V_G$ goes further to the negative values, the slope flattens but does not fully saturate. Finally, the conductance approaches the initial CNP value as $V_G$ returns to the origin. As the $V_G$ sweep range increases, the distance between the two CNPs and the saturated $I_{SD}$ region increase systematically.

The hysteresis behaviors were observed in different graphene-TMO devices and attributed to either the ferroelectricity of the TMO layer or charge trapping at the graphene-TMO interface. First, spontaneous polarization from a ferroelectric layer beneath graphene can introduce hysteretic conductance by directly modulating the induced electric field.[16-17,19,29] STO might exhibit ferroelectricity as well if its original cubic structure changes to a tetragonal structure because of either epitaxial strain or cation off-stoichiometry.[30-31] It has also been reported that local displacement of oxygen ions can induce ferroelectric-like effect.[32] However, our homoepitaxial STO thin film grown at high oxygen partial pressure is free from strain and stoichiometric, indicating that it is not a ferroelectric macroscopically (see the XRD results in Figure 1a).[7] It should also be noted that the direction of hysteresis in our device (*i.e.*, anti-hysteresis) is the opposite to those observed in graphene on ferroelectric layers, which result from alignment of spontaneous dipole moments.[17,19,32] Second, charge trapping at the graphene-TMO interface can induce the hysteretic conductance.[33-35] Indeed, most studies attribute extrinsic defects or unintentionally adsorbed water molecules at the interface to charge trapping sites.[23,36-37]

In our high-quality device, oxygen vacancies, instead of extrinsic defects at the graphene-STO interface, are considered as dynamic charge trapping sites. As mentioned above, oxygen vacancies can be introduced in the STO layer near the metal-STO heterojunctions by electroforming.[10-11] The preferential redox activity near the conducting electrode and the apparent migration of oxygen vacancies due to the applied electric field further support the assertion that oxygen vacancies should be accumulated near the interface.[4] Based on the dynamics of oxygen vacancies, which is schematically expressed in Figure 2b, hysteretic conductance in our device can be understood. As positive $V_G$ is applied up to $V_{th}$, a conventional field-dependent charge carrier injection into the graphene layer is expected (Figure 2b A). When the $V_G$ sweep exceeds $V_{th}$, the conductance begins to saturate and the hysteresis develops owing to the electroforming of oxygen vacancies at the interface between the graphene and STO (Figure 2b B).[1,38] The higher the applied $V_G$, the greater the number of oxygen vacancies that accumulate. The field-induced generation of charge trap sites prevents excess electrons from being injected into the graphene, resulting in the saturated conductance. Merely decreasing $V_G$ does not annihilate electroformed oxygen vacancies, as shown in Figure 2b C. Therefore, the electron concentration in the graphene decreases immediately as $V_G$ decreases. The instantaneous reduction of the carrier density in the graphene naturally introduces the shift of the CNP to the right. As $V_G$ further decreases and becomes negative, oxygen vacancies start to disappear and usual hole injection into the graphene begins. (Figure 2b D) When oxygen vacancies are fully annihilated at large negative $V_G$, the conductance curves return to the original CNP values as illustrated in Figures 2b E and F.

By applying a magnetic field, quantum Hall states are clearly observed together with the hysteretic conductance, as shown in **Figures 3**a and 3b. The observation of the quantum Hall effect indicates the robustness and high sensitivity of the graphene probe. It also supports that the intrinsic electronic property of the graphene is maintained while the modification (hysteresis) originates from the dynamic changes in the dielectric environment underneath the graphene. In particular, the observation of quantum Hall behavior indicates that our device is mostly free from extrinsic defects that might act as charge-trapping sites. Moreover, extrinsic charge-trapping sites would induce hysteretic conductance even at the lowest $V_G$.[39]

So far, quantum Hall states in graphene-STO device were only observed by using either bulk single-crystalline or high-quality epitaxial single-crystalline thin film STO.[27-28] However, we also note that those studies did not observe any hysteretic quantum conductance. The electric fields estimated from applied $V_G$ and the thickness of the STO layers are 0.3 and 33 kV/cm for the bulk (500 μm) and thin film (300 nm) cases, respectively. These values are considerably smaller than the applied electric field in our current device with a film thickness of ~90 nm, which was 110 kV/cm for $V_G = 1$ V. Therefore, we can conclude that a certain amount of electric field (equivalent to $V_G = V_{th}$ in our device) is necessary for the electroforming of oxygen vacancies and opening of the hysteresis in the quantum conductance.

The quantum Hall behavior provides further information regarding the carrier responses within a dynamic dielectric environment. Systematic tendencies in the conductance behavior, i.e., a *proportional* expansion of the width of the conductance plateau and an increase in the distance between the two CNPs, are observed when the $V_G$ sweep range is increased (Figures 3a and 3b). Upon aligning the CNPs, *i.e.*, subtracting $V_{CNP}$s from $V_G$s, and multiplying a scaling parameter *α* to ($V_G − V_{CNP}$), we could converge all the curves together as shown in Figures. 3c and 3d. In other words, the curves could be well scaled across a large $V_G$ sweep range regardless of the application of the magnetic field, especially for a negative $V_G$ sweep. The deviation observed for a positive $V_G$ sweep suggests non-linear and/or incomplete annihilation of oxygen vacancies at large negative $V_G$ range. The universal scaling of the quantum conductance convinces us to consider a change in the effective capacitance in terms of the $V_G$ sweep range ($V_{G\text{-range}}$) in our device.

Upon analyzing the physical meaning of the scaling parameter $\alpha$, a clear *linear* behavior was observed between $1/\alpha$ and the applied $V_{G\text{-range}}$, as shown in **Figure 4**a. The value of $\alpha$ at a specific $V_{G\text{-range}}$, therefore, follows the relation,

$$\frac{1}{\alpha} - 1 = b\left(V_{G-\text{range}} - V_{\text{th}}\right), \tag{1}$$

where $b$ is the slope. From Figure 4a, the $V_{\text{th}}$ value can be more precisely defined as ~0.75 V. It is worthwhile to note that the above relation only holds when $V_{G\text{-range}} \geq V_{\text{th}}$. It means that the conductance hysteresis does not occur when the $V_G$ scans in the narrow range (i.e. $V_{G\text{-range}} < V_{\text{th}}$), and the change in the dielectric environment starts to occur when $V_{G\text{-range}}$ exceeds $V_{\text{th}}$. Since the conductance change in Figure 3 is driven by the carrier density modulation in the graphene due to $V_{G\text{-range}}$, its scaling by $\alpha$ reflects the strength of effective electric field applied to graphene. Furthermore, considering the induced charge carrier density in graphene, the effective gate-voltage, *i.e.* $\alpha(V_G - V_{\text{CNP}})$, corresponds to the change in the capacitance of dielectric layer. This can be alternatively expressed as

$$C_{\text{tot}} = \alpha C_{\text{STO}}, \tag{2}$$

where $C_{\text{tot}}$ and $C_{\text{STO}}$ correspond to the total capacitance of the gate-dielectric layer showing the hysteretic conductance behavior and the capacitance of the STO layer within a non-hysteretic condition, i.e., without oxygen vacancies, respectively.

Specifically considering the dynamics of oxygen vacancy inside STO illustrated in Figure 2b, we constructed a simple layer model of our graphene-STO device, as depicted in Figure 4b. The model consists of a graphene top electrode, an STO thin film of total thickness $d$ and dielectric constant $\varepsilon_{STO}$, and an Nb:STO bottom electrode. In order to take into account the influence of oxygen vacancies on the dielectric properties of STO, we considered a thin STO layer with oxygen vacancies (OV layer) having the capacitance $C_{OV}$ just below the graphene layer. The OV layer was simplified to have a uniform dielectric constant $\varepsilon_{OV}$ ($\leq \varepsilon_{STO}$), and we estimated that it would increase in thickness $d_{OV}$ with increasing oxygen vacancy concentration ($d_{STO} + d_{OV} = d = 90$ nm). Then the total capacitance ($C_{tot}$ with the dielectric constant $\varepsilon_{tot}$) coming from the serial connection of two capacitors, $C_{OV}$ and $C_{STO}$ shows the relation $1/C_{tot} = 1/C_{OV} + 1/C_{STO}$, which gives the following expression:

$$\frac{d}{\varepsilon_{tot}} = \frac{d_{OV}}{\varepsilon_{OV}} + \frac{d - d_{OV}}{\varepsilon_{STO}}. \tag{3}$$

From Equations. 2 and 3, we obtain

$$\frac{d_{OV}}{\varepsilon_{OV}} - \frac{d_{OV}}{\varepsilon_{STO}} = \frac{d}{\varepsilon_{STO}}\left(\frac{1}{\alpha} - 1\right). \tag{4}$$

From Equations 1 and 4,

$$d_{OV}\left(\frac{1}{\varepsilon_{OV}} - \frac{1}{\varepsilon_{STO}}\right) \propto \left(V_{G-range} - V_{th}\right), \tag{5}$$

where the left term expresses the conversion of normal STO layer to an oxygen-vacant STO layer. Equation 5 indicates that generation of OV layer with thickness $d_{OV}$ and dielectric constant $\varepsilon_{OV}$ is driven by the excessive gate-voltage applied to the total STO layer above $V_{th}$. Here, the variation of $d_{OV}$ and that of $\varepsilon_{OV}$ cannot be extracted separately because they are coupled in one equation. It implies that the increment of $V_{G-range}$ above $V_{th}$ would increase $d_{OV}$ and/or decrease $\varepsilon_{OV}$ based on the oxygen vacancy accumulation.

Since $\varepsilon_{STO}$ can be estimated from the universal quantum plateaus in Figure 3c,[28] a quantitative relation among $d_{OV}$, $\varepsilon_{OV}$, and $V_{G-range}$ can be obtained from Equation 5. Figure 4c presents the evolution of $d_{OV}$ and $\varepsilon_{OV}$ as a function of $V_{G\text{-range}}$. With increasing $V_{G\text{-range}}$, both the increase of the thickness and the decrease of the dielectric constant of the OV layer can be explained in terms of the creation of oxygen vacancies. More importantly, if either $d_{OV}$ or $\varepsilon_{OV}$ can be precisely determined (for example, $d_{OV}$ can be obtained from *in-situ* TEM[40]), one can completely understand the quantitative dielectric nature of the STO layer with dynamic oxygen vacancy accumulation.

It should be emphasized that the graphene probe enabled the observation above. In addition to the quantum transport behavior described above, the van der Waals bonding nature and structural integrity of graphene effectively prevent the graphene layer from chemically interacting with the layer underneath on the atomic level. This is not the case for most conventional electrodes in use. For example, typical capacitance measurements might suffer from strong chemical interaction between a metal electrode and an oxide layer, especially in the circumstances of redox activity. Indeed, oxidation of the metal electrode itself during the redox reaction might obscure the understanding of the dielectric properties of the oxide layer.[41-42] In addition, usual AC capacitance measurements should take into account time-dependent electroforming and dynamic movement of oxygen vacancies, while our measurement can be more simply analyzed. Owing to the usefulness of our approach, we could fully account for accumulated oxygen vacancies at a certain $V_G$, through the shape of the conductance curves. Therefore, the quantum Hall probe can be a useful method for the investigation of the field-induced electroforming process in transition metal oxides.

In summary, we investigated electroforming of oxygen vacancies in a STO thin film using the sensitive carrier response of graphene. The hysteretic quantum Hall conductance was understood in terms of the creation and annihilation of oxygen vacancies at the interface between graphene and a STO thin film. Experimentally, a linear scaling was observed between the capacitance and the sweep range of $V_G$, which led to the quantitative relation between the thickness and dielectric constant of the STO layer with oxygen vacancies. We believe that a quantum conductance probe using graphene can be utilized to understand various electronic and ionic behaviors in different transition metal oxides.

**Experimental Section**

*Thin film growth and structural characterization.* Epitaxial STO thin films were grown on atomically flat (001)-oriented single crystalline Nb (0.5wt%) doped STO substrates (Nb:STO) using pulsed laser epitaxy (PLE) at 700 °C. An excimer (KrF) laser with a wavelength of 248 nm (IPEX 864, Lightmachinery, Nepean, Canada), an energy fluence of 0.8 Jcm$^{-2}$, and a repetition rate of 5 Hz was used. In order to minimize the leakage current owing to possible defects of the thin films, we used high oxygen partial pressure atmosphere of 100 mTorr for the growth and subsequently post-annealed the samples at 400 °C in 400 Torr of oxygen for 1 h. The atomic structures and the crystallinity of the epitaxial thin film were examined using high-resolution x-ray diffraction (XRD). By using the x-ray reflectivity, the thickness of the thin film was determined as 90 nm.

*Graphene based transistor fabrication.* A graphene flake was prepared using mechanical exfoliation on a poly(methyl methacrylate) (PMMA)-coated silicon substrate, and Raman spectroscopy confirmed that it was a monolayer. The exfoliated graphene flake was transferred onto the STO//Nb:STO with the dry-transfer method, and the graphene field effect transistor with thermally evaporated Cr/Au (5/50 nm) electrodes was fabricated using a conventional electron-beam lithography technique. The distance between source and drain was 2 μm.

*Transport measurement.* The electrical transport experiments were conducted at the 2 K under magnetic fields of 0 and 14 T using a cryostat (PPMS-Dynacool, Quantum Design Inc.) and a semiconductor parameter analyzer (B1500A, Agilent Technologies). All of the measurements are performed under the 1 mV for the source-drain voltage with the condition of "high-vacuum mode" with the scan rate of 10 mV per 0.5 sec.

**Supporting Information**

Supporting Information is available from the Wiley Online Library or from the author.

**Acknowledgements**

K.T.K., H.K., and J.P. contributed equally to this work. We appreciate S. Lee for helpful discussion. This work was supported by the Basic Science Research Programs ((NRF-2014R1A2A2A01006478 (W.S.C.) and NRF-2015R1D1A1A01059850 (H.K.)) through the National Research Foundation of Korea.

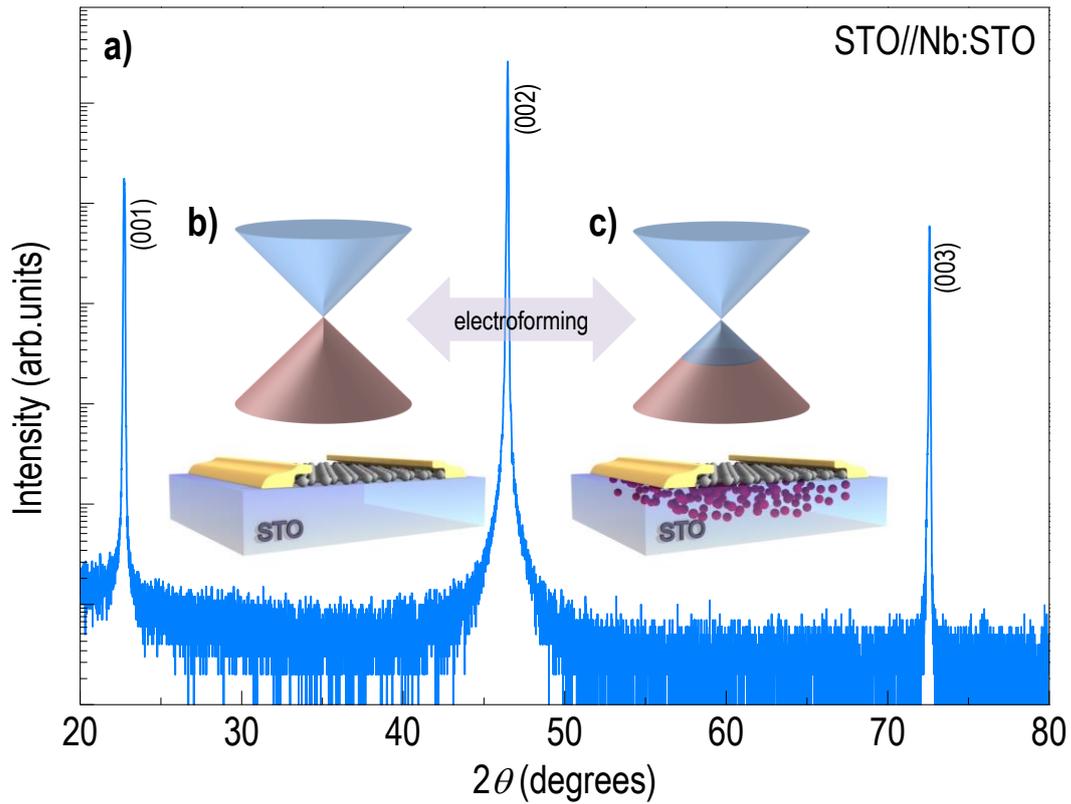

**Figure 1.** Observation of electroforming of oxygen vacancies at the high-quality graphene-STO interface. a) The x-ray diffraction $\theta$-$2\theta$ scan of a single-crystalline epitaxial STO thin film on a (001) Nb-doped STO substrate. b),c) The schematic diagrams of our graphene-STO heterostructure devices. As oxygen vacancies accumulate at the interface between the graphene and STO, hole-doped behavior in the graphene is expected.

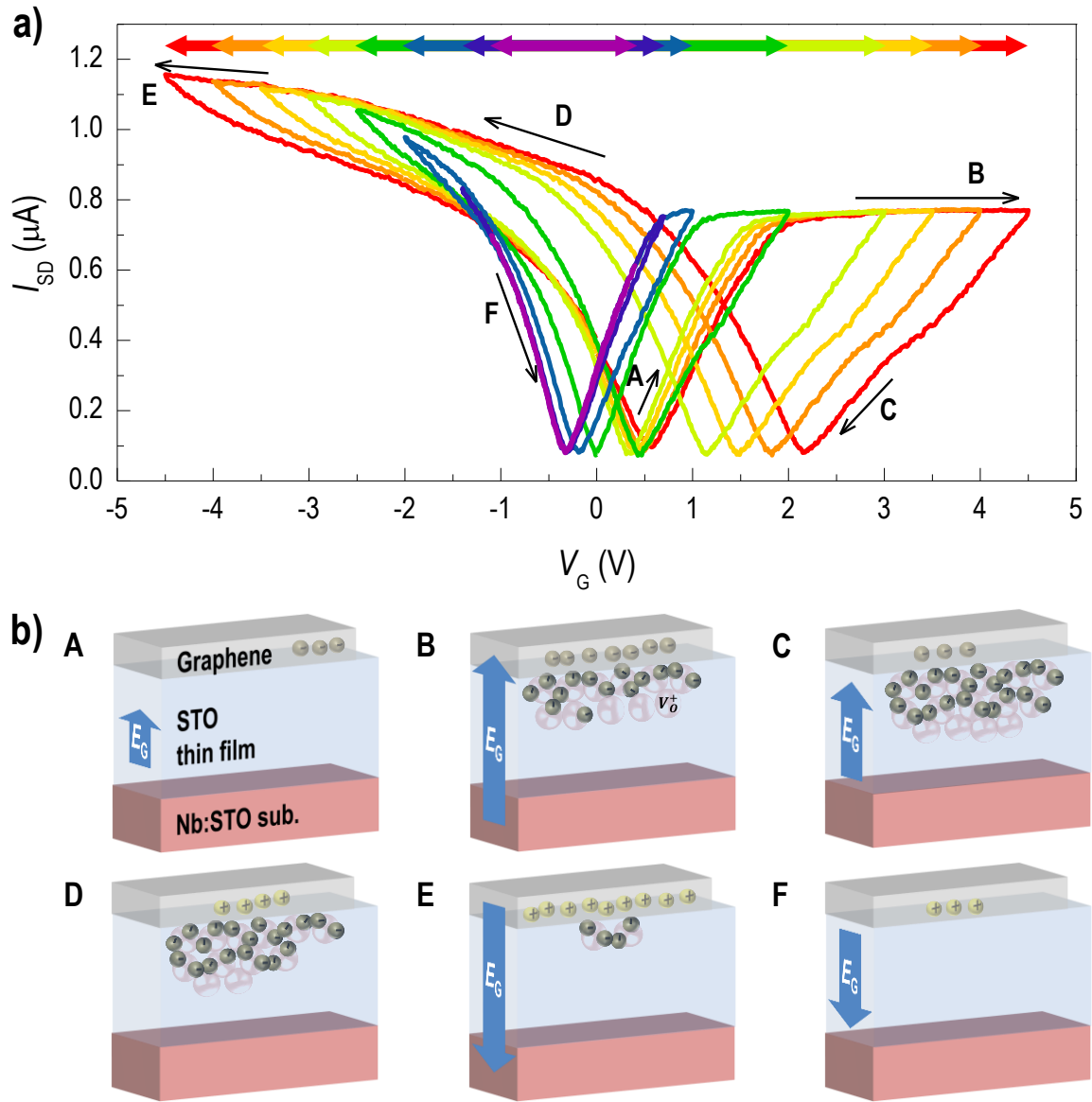

**Figure 2.** Hysteretic conductance behavior in the graphene-STO device. a) The source-drain current as a function of the gate voltage at 2 K. The $V_G$ sweep range was systematically expanded from –1, +0.5 V (purple loop) to ±4.5 V (red loop). The arrows and alphabets (A–F) indicate the $V_G$ sweep sequence. The $V_G$ sweep started from 0 to positive $V_G$ and to negative $V_G$ and then back to 0. b) The schematic diagrams representing the creation and annihilation of oxygen vacancies. The asymmetric creation and annihilation of oxygen vacancies and resultant trap levels are responsible for the hysteretic behavior.

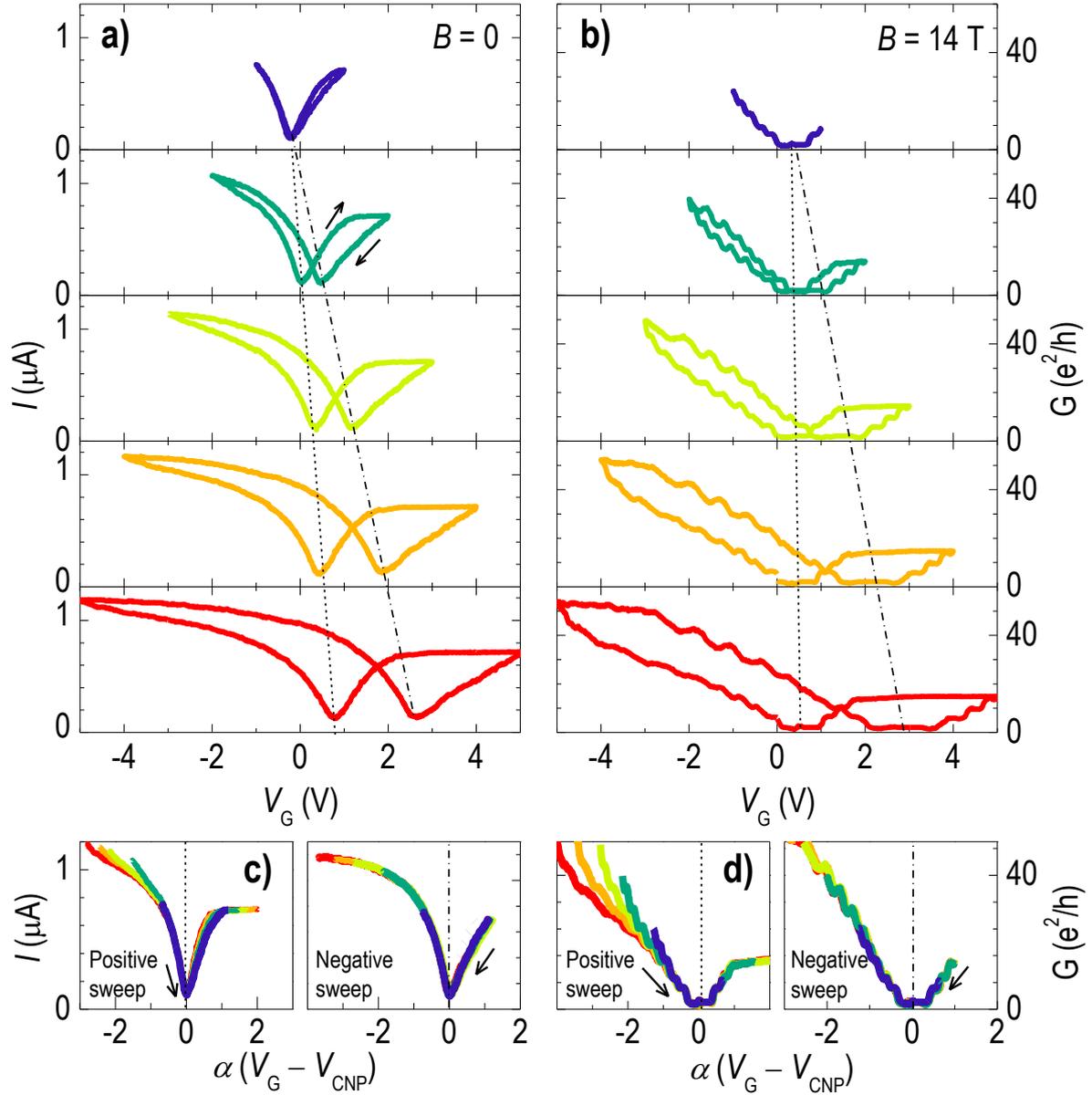

**Figure 3.** Evolution of the hysteretic quantum Hall conductance due to oxygen vacancy creation. The conductance a) without and b) with a magnetic field (14 T) under the $V_G$ sweep range of ±1, ±2, ±3, ±4 and ±5 V at 2 K. The vertical guidelines indicate a linear shift of charge neutrality points with respect to the $V_G$ sweep range. Aligning charge neutrality points by subtracting $V_{CNP}$ from $V_G$ and scaling the applied voltage by multiplying a scaling factor $\alpha$, at $B =$ c) 0 and d) 14 T. The grey dashed curves indicate the conductance in the low-voltage (<1 V) region.

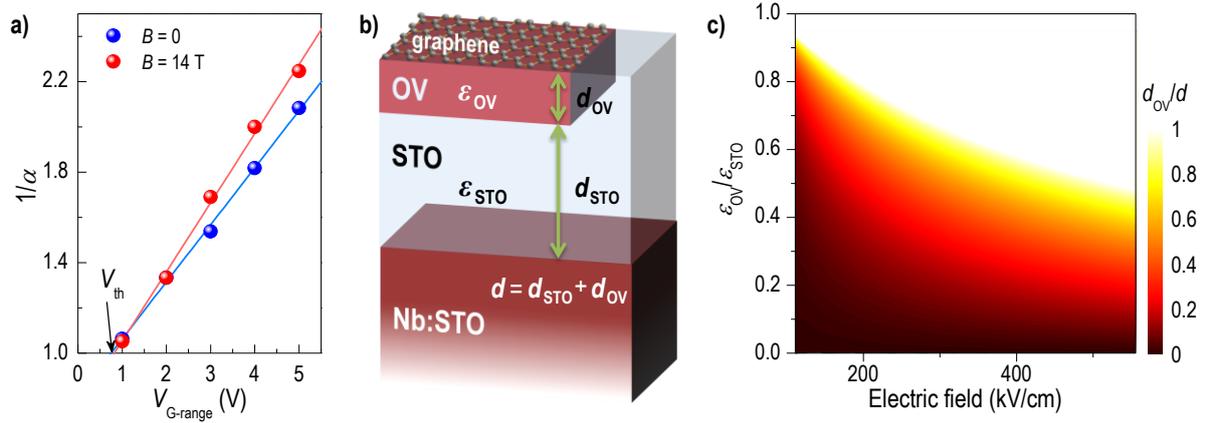

**Figure 4.** Dielectric properties of the STO thin film with oxygen vacancies. a) The $1/\alpha$ value as a function of $V_{\text{G-range}}$ showing a clear linear dependence. $V_{\text{th}}$ is nearly the same, approximately 0.75 V, independent of the applied magnetic field. b) A capacitor model including the oxygen vacancy layer is adopted in order to understand the experimental data. c) The color mapping of $d_{\text{OV}}$ as functions of $\varepsilon_{\text{OV}}$ and $V_{\text{G-range}}$, describing the range of thickness and dielectric constant of the STO thin film with oxygen vacancies.

Supporting Information

**Quantum Conductance Probing of Oxygen Vacancies in SrTiO$_3$ Epitaxial Thin Film Using Graphene**

*Kyeong Tae Kang, Haeyong Kang, Jeongmin Park, Dongseok Suh\*, and Woo Seok Choi\**

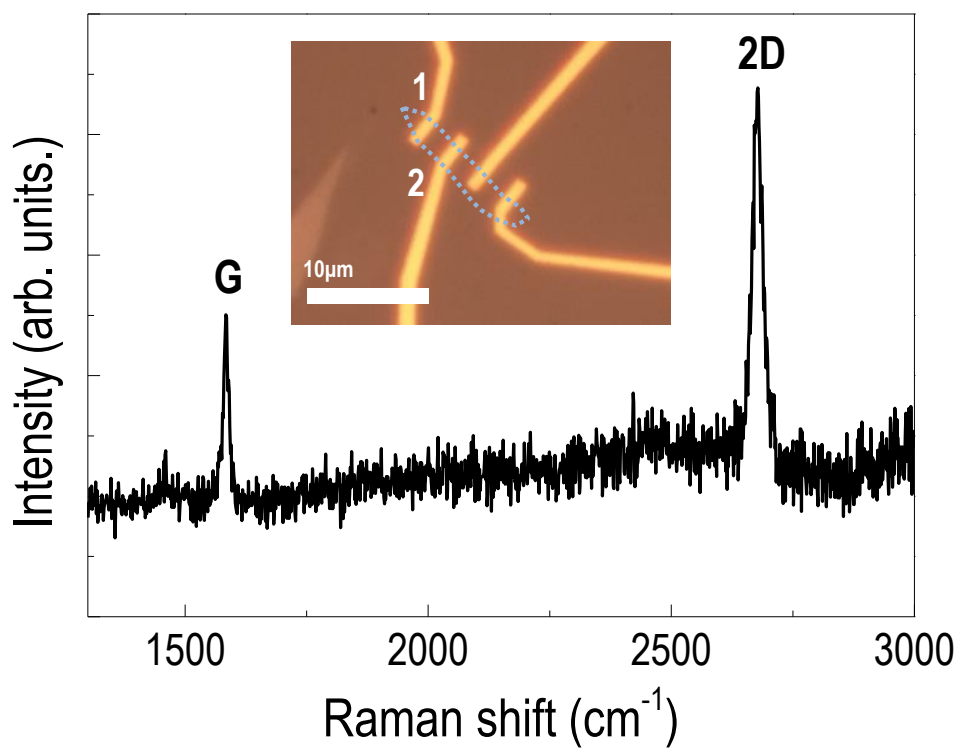

**Figure S1.** Raman spectrum of the exfoliated monolayer graphene flake. The inset is the fabricated device observed through optical microscopy. 1 and 2 represents the source and drain electrodes, respectively.

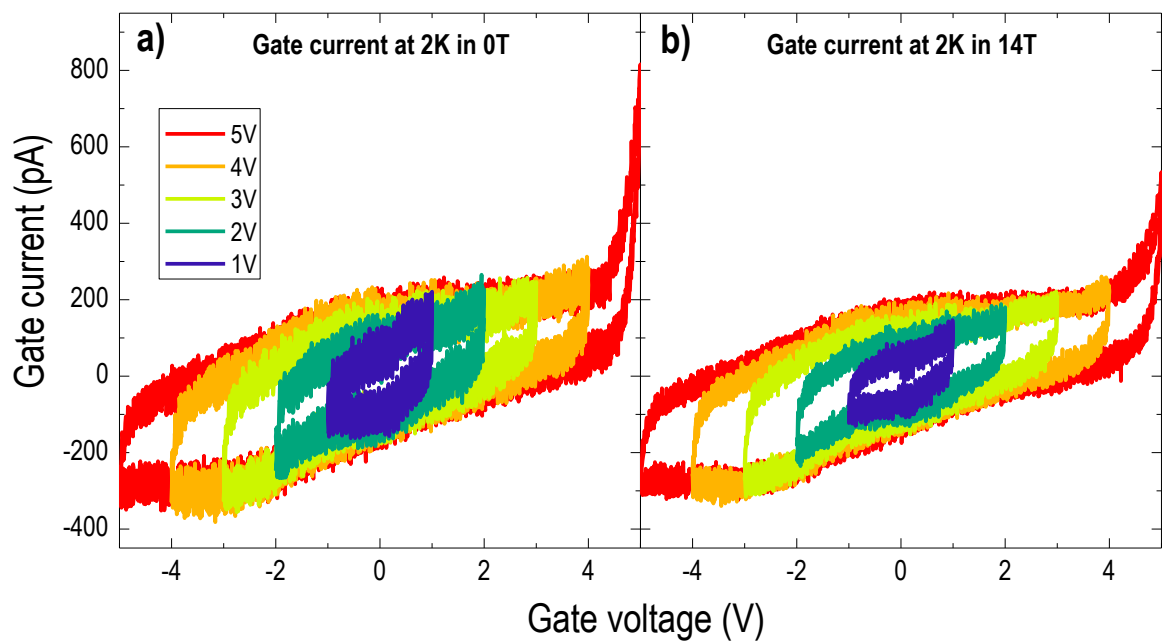

**Figure S2.** Gate leakage current measurements a) without b) with a magnetic field (14 T), during the measurement presented in Figure 3.